\documentclass[%
	reprint,
	superscriptaddress,
	amsmath,amssymb,
	pra,
]{revtex4-2}

\usepackage[T1]{fontenc}
\usepackage{babel}

\usepackage{tensor}

\usepackage{hyperref}
\usepackage{orcidlink}

\usepackage{graphicx} 
\usepackage{wrapfig}
\usepackage{caption}
\usepackage{subcaption}
\usepackage{svg}

\usepackage{amsmath} 
\usepackage{amssymb}

\usepackage[normalem]{ulem}

\usepackage[toc,page]{appendix}

\usepackage{parskip}

\usepackage{xcolor}

\newcommand{\ket}[1]{\left|{#1}\right\rangle}

\begin{document}
	
    \preprint{APS/123-QED}
    
    \title{Toroidal Transitions in Hydrogenic and Alkali Atoms}
    
    %%%%%%%%%%%%

    \author{Kai Xiang Lee \orcidlink{0009-0008-7767-4755}}
    \affiliation{Centre for Quantum Technologies, Nanyang Technological University, Singapore, Singapore.}
    
    \author{Vincent Mancois \orcidlink{0000-0001-9441-7461}}
    \affiliation{Centre for Disruptive Photonic Technologies, SPMS, The Photonics Institute, Nanyang Technological University, Singapore 637371, Singapore.}
    \affiliation{MajuLab, International Joint Research Unit IRL 3654, CNRS, Université Côte d’Azur, Sorbonne Université, National University of Singapore, Nanyang Technological University, Singapore, Singapore.}
    
    \author{Kelvin Lim \orcidlink{0000-0002-8356-050X}}
    \affiliation{Centre for Disruptive Photonic Technologies, SPMS, The Photonics Institute, Nanyang Technological University, Singapore 637371, Singapore.}
    
    \author{David Wilkowski \orcidlink{0000-0002-4712-9456}}
    \affiliation{Centre for Quantum Technologies, National University of Singapore, Singapore 117543, Singapore.}
    \affiliation{Centre for Disruptive Photonic Technologies, SPMS, The Photonics Institute, Nanyang Technological University, Singapore 637371, Singapore.}
    \affiliation{MajuLab, International Joint Research Unit IRL 3654, CNRS, Université Côte d’Azur, Sorbonne Université, National University of Singapore, Nanyang Technological University, Singapore, Singapore.}
    
    %%%%%%%%%%%%%%%%%5
    
    \date{\today}
    
    \begin{abstract}
        In addition to electric and magnetic, expansion of current density also yields toroidal, a lesser-known family of multipoles. A recent proposal, I. Kuprov \textit{et al.}, Science Adv. 8 abq6751 (2022), explores the possibility of a direct observation of optical toroidal transitions in hydrogen and alkali atoms in the presence of a large magnetic field that decouples the spin and the angular momentum of the electron. However, the difficulty of observing these transitions against the nearby electric dipole (E1) transitions were underestimated because of extra admixture coming from diamagnetic coupling. Here, we revisit the toroidal coupling in atoms, taking diamagnetic contribution into account, and discuss the technical challenges of observing toroidal coupling in atomic physics. We show that toroidal transition should be searched in transitions with low principal quantum numbers. The remaining strong electric-dipole contribution could be removed using an appropriate differential measurement. 
    \end{abstract}
    
    \maketitle
    
    %%%%%%%%%%%%%%%%%%%%%%%%%%%%%%%%%%%%%%%%%%%%%%%

\section{Introduction}

Since the first conceptualization of the toroidal dipole by Ia. B. Zel’dovich in 1957, there have been many studies on toroidal structures in atoms \cite{Zeldovich1957}. Most famously, the measurement of nuclear toroidal moment has found applications as a probe to constrain parity non-conserving (PNC) interactions in the standard model \cite{Wood1997, Gomez2007, Sheng2010, Gwinner2022, Damitz2024, Tsigutkin2009}. Beyond PNC measurements, literature on toroid structures in atoms remains sparse. Studies proposed hydrogenic toroid polarizabilities \cite{Costescu1991}, Stark-induced anapole magnetizability and polarizabilities \cite{Lewis1995, Lewis1998, Bhattacharya2000, Mielewczyk2006, Summa2022}, anapole-controlled spontaneous emission \cite{ZuritaSanchez2021}, toroid multipolar Rayleigh scattering \cite{Costescu1994}, electroweak nuclear-induced anapoles \cite{Boston1990, Apenko1982, Bhattacharya1995}, inter-electronic anapole moments \cite{Lewis1993}, and collective toroidal excitations in atom arrays \cite{Ballantine2020}. \par

One recent study describes direct electronic excitation under toroidal dipolar transitions, via the magneto-electric relativistic response \cite{Kuprov2022}. In this paper, we aim to provide a further comprehensive analysis of the toroidal dipole transition in hydrogenic atoms. \par

The paper is organized as follows: In Section II, we remind the general properties of toroidal excitations in comparison with the standard magnetic and electric ones. In section III, starting from the Dirac equation, we derive the Hamiltonian and explain the different approximations. In section IV, we compute the relative weight of toroidal and electric dipole contributions. As with diamagnetic interaction, the electric coupling always dominates; we introduce a differential method, leveraging selection rules, to extract the weaker toroidal contribution. Finally, the work is concluded and put in perspective in section V.
    
    %%%%%%%%%%%%%%%%%%%%%%%%%%%%%%%%%%%%%%%%%%%%%%%

\section{Electronic Toroidal Transitions}
\label{sec:toroidal_multipole_properties}

Under a static distribution of charges and currents, the electric and magnetic multipole families form a complete basis for electromagnetic field decomposition. However, when the sources exhibit time-dependence, an additional family of multipoles appear: the toroidal multipoles \cite{Dubovik1990, Dubovik1986, Nanz2016}. They can be distinguished from the electric and magnetic families by their symmetry properties under space and time inversion (\autoref{tab:multipole_symmetry_properties}). Here, the Electric Toroidal gives the weakest contribution and, thus, will be disregarded in the rest of the paper. Then, the magnetic toroidal is renamed in short "toroidal" \par

\begin{figure}
    \centering
    \includegraphics[width=\linewidth]{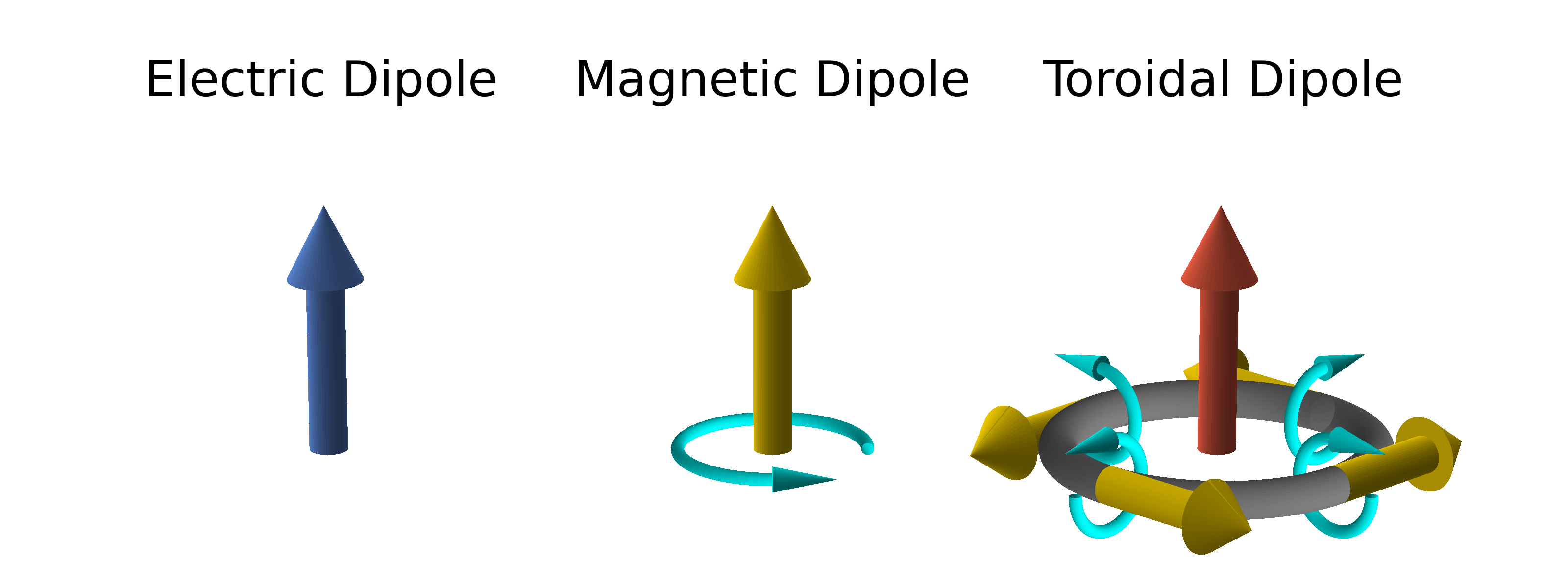}
	\begin{tabular}{ |l|c|c|c| }
		\hline
		Transition & $\Delta \ell$ & $\Delta m_\ell$ & $\Delta m_s$ \\ 
		\hline
		Electric Dipole & $\pm1$ & $0,\pm1$ & $0$      \\
		Magnetic Dipole & $0$    & $0,\pm1$ & $0,\pm1$ \\
		Toroidal Dipole & $\pm1$ & $0,\pm1$ & $0,\pm1$ \\
		\hline 
	\end{tabular}
    \caption{Dipoles from each multipole family, along with associated selection rules in the Paschen-Beck regime}
    \label{fig:electronic_dipoles}
\end{figure}

In atoms, dynamic multipoles can be defined on electronic transitions - the matrix element between any two states giving a non-zero contributions. This property is reflected in the selection rules of a transition, as illustrated for the dipole moments in \autoref{fig:electronic_dipoles}. As the latter are the lowest ($k=1$) order terms, they give the strongest contributions. Hence, in this work, we disregard the higher order terms.  \par

\begin{table}[t]
	\centering
	\caption{Symmetry parity properties of the $k$th-multipole under spatial and temporal inversion}
	\begin{tabular}{ |c|c|c| }
		\hline
		Multipole Family & Spatial Parity $\mathcal{P}$ & Temporal Parity $\mathcal{T}$ \\ 
		\hline
		Electric & $(-1)^k$ & $+1$ \\  
		Magnetic & $-(-1)^{k}$ & $-1$ \\ 
		(Magnetic) toroidal & $(-1)^k$ & $-1$ \\  
		Electric toroidal & $-(-1)^k$ & $+1$ \\  
		\hline 
	\end{tabular}
	\label{tab:multipole_symmetry_properties}
\end{table}

Toroidal transitions have the same spatial parity as electric transitions, and so any electric-allowed transitions between orbitals will have a corresponding toroidal-allowed transition. At the same time, toroidal transitions have the same temporal parity as magnetic transitions. For the dipole, it allows for a spin-flip otherwise forbidden under electric transitions. These two properties in conjunction allow for a spin toroidal dipole transition to be uniquely distinguished from the well-studied electric and magnetic transitions \cite{Kuprov2022}.  \par

The spin toroidal dipole (T1) transition arises from relativistic corrections to the Schrodinger equation. It first appears in the angular magnetoelectric part of the spin-orbit term \cite{Kuprov2022}, or equivalently, in the second-order retardation correction of the Dirac's equation \cite{Marian1996} (see \autoref{Appendix}). As they arise from higher-order terms of the expansion, the toroidal dipole moment is much weaker than the electric dipole (E1) or magnetic dipole (M1) moments. \par

\begin{figure}[t]
    \centering
    \includesvg[width=\linewidth]{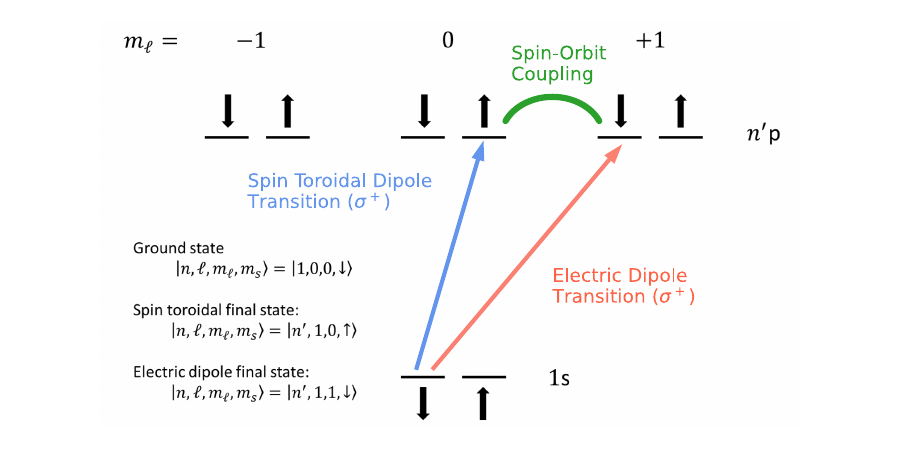}
    \caption{Electric dipole (blue) and spin toroidal dipole (red) transitions from the ground state for the hydrogen atom under right-handed circularly-polarized illumination ($1\mathrm{s}\to n'\mathrm{p}$). The $m_\ell,m_s$ states are mixed by the spin-orbit coupling (green).}
    \label{fig:LevelDiagram}
\end{figure}

In a hydrogen-like atom, the spin-orbit coupling mixes the orbital and spin angular momenta, such that $m_\ell$ and $m_s$ are no longer good quantum numbers. In the total angular momentum $j,m_j$ basis, all T1 transitions have spectral collisions with E1 transitions (\autoref{fig:LevelDiagram}), with a factor of $(\alpha a_0 k)^{-2} \sim 10^{11}$ stronger, which makes spectroscopic observation challenging. To improve the T1/E1 transition strength ratio, it was proposed to isolate the spin T1 transition by decoupling the orbital and spin angular momentum applying a large magnetic field. In addition, as the spin-orbit coupling scales as $\propto1/r^3\sim n^{-3}$, it is expected that a ground state to Rydberg state transition would be favorable. An estimate suggested that the T1 transition rate exceeds that E1 transition rate with magnetic field $B_0 \sim 5\ \mathrm{T}$ for the hydrogen Balmer series $n=2\to n'=51$ transition \cite{Kuprov2022}. However, this estimate does not take into account the diamagnetic level-mixing effects that dominate interactions in large magnetic fields \cite{Gallagher1994,Courtney_1996_LiChaos, Friedrich1989}. \par

We study the T1 transition in more detail, taking into account these level-mixing effects. Because eigenstates of different $\ell$ are well-separated, we only consider the visibility of the T1 transition against the E1 transition, being primarily mixed by the fine-structure spin-orbit term. The M1 transitions are far-detuned and can be neglected. \par
    %%%%%%%%%%%%%%%%%%%%%%%%%%%%%%%%%%%%%%%%%%%%%%%

\section{Computing Transition Rates}
\label{sec:calculate_transition_rates}

The matrix elements are computed between states in the uncoupled basis $\alpha=n,\ell,m_\ell,m_s$, where the electronic wavefunction can be separated into radial, angular, and spin components, $\psi(\alpha; r,\theta,\phi) = R_{n\ell}(r) Y_{\ell m_\ell}(\theta,\phi) S_{m_s}$. We assume that the electromagnetic field does not affect the nuclear spin, and only consider the electronic degrees of freedom. In this representation, the matrix elements for a given operator $\mathcal{O}$ can be decomposed into

\begin{equation}
    \langle \alpha^f | \mathcal{O} | \alpha^i \rangle
    =
    \mathcal{R}^k_{n^i\ell^i\to n^f\ell^f}
    \mathcal{A}^{k,q}_{\ell^i m_\ell^i\to \ell^f m_\ell^f}
    \mathcal{S}_{m_s^i\to m_s^f}
\end{equation}

Here, the initial and final states are labeled with superscripts. The radial integrals are evaluated using recurrence relations \cite{Sanchez1992} and validated against standard numerical methods as well as known analytic solutions \cite{WolframLG}. The angular matrix elements are computed with the Gaunt coefficients, and the spin part can be evaluated through the Pauli matrix elements between states. \par

\begin{align}
    \mathcal{R}^k_{n^i\ell^i\to n^f\ell^f}
    &=
    \int 
    R_{n^i\ell^i}(r) r^k R_{n^f\ell^f}(r)
    \ r^2\text{d}r
    \label{eq:Radial_Overlap_Integral}
    \\
    \mathcal{A}^{k,q}_{\ell_i m_\ell^i\to \ell_f m_\ell^f}
    &=
    \int
        Y_{\ell^f m_\ell^f}(\Omega)^{*}
        Y_{kq}(\Omega)
        Y_{\ell^i m_\ell^i}(\Omega)
    \ \text{d}\Omega
\end{align}

We consider an electron bound to a hydrogenic potential of charge $+Ze$ in a static magnetic field $\boldsymbol{B_0} = B_0 \hat{e}_z$, although the methods described here can be extended to other alkali atoms \cite{Weber2017}. We work in the Coulomb gauge. The atom is illuminated by an optical driving field along the magnetic field axis with circular polarization, $\boldsymbol{E_0}=\frac{E_0}{\sqrt 2} 
    e^{ikz}
    \left(
    e^{-i\omega t}\hat \epsilon_\pm
    -
    e^{i\omega t}\hat \epsilon_\mp
\right)$. The total electromagnetic environment seen by the electron are given by the electric and magnetic fields,

\begin{eqnarray}
    \boldsymbol E (\boldsymbol r, t)
    &&=
    \frac{Ze}{4\pi\epsilon_0} 
    \frac{\boldsymbol r}{|r|^3}
    -
    \frac{E_0}{\sqrt 2} 
    e^{ikz}
    \left(
    e^{-i\omega t}\hat \epsilon_\pm
    -
    e^{i\omega t}\hat \epsilon_\mp
    \right)
    \\
    \boldsymbol B (\boldsymbol r, t)
    &&=
    -
    \frac{i}{\sqrt 2} \frac{E_0}{c} 
    e^{ikz}
    \left(
    e^{-i\omega t}\hat \epsilon_\mp
    +
    e^{i\omega t}\hat \epsilon_\pm
    \right)
    -
    B_0 \hat{e}_z
\end{eqnarray} 

We begin with the second-order Foldy-Wouthuysen expansion of the Dirac Hamiltonian \cite{Foldy1950}. Because the spin-toroidal term emerges at this order, it is sufficient to truncate the relativistic expansion up to terms of this order \cite{Mondal2015,Kuprov2022}. The total Hamiltonian is thus

\begin{equation}
    H = H_0 + H_\text{FS}
\end{equation}

The minimal-coupling hydrogen Hamiltonian is given by $H_0$, and the fine-structure terms are contained in $H_\text{FS}$, 

\begin{align}
    H_0 
    &=
    \frac{(\boldsymbol{p}+e\boldsymbol{A})^2}{2m}
    - e\Phi
    + \frac{e\hbar}{2m} \boldsymbol{\sigma\cdot B}
    \\
    H_\text{FS}
    &=
    - \frac{(\boldsymbol{p}+e\boldsymbol{A})^4}{8m^3c^2}
    + \frac{e\hbar^2}{8m^2c^2} \nabla\cdot \boldsymbol{E}
    \nonumber
    \\
    &\ \ \ 
    +
    \frac{e\hbar}{8m^2c^2}\boldsymbol{\sigma}\cdot
    \left(
    \boldsymbol{E}
    \times(\boldsymbol{p}+e\boldsymbol{A})
    -
    (\boldsymbol{p}+e\boldsymbol{A})\times\boldsymbol{E}
    \right)
\end{align}

Here, we rewrite some of the electromagnetic terms in the form of the scalar potential $\Phi$ and vector potential $\boldsymbol{A}$, where $\boldsymbol{B} = \boldsymbol{\nabla}\times\boldsymbol{A}$ and $\boldsymbol{E} = \partial\boldsymbol{A}/\partial t - \boldsymbol{\nabla}\Phi$.

\begin{align}
    \Phi
    &=
    \frac{Ze}{4\pi\epsilon_0}\frac{1}{r}
    \\
    \boldsymbol A
    &
    = 
    \frac{i}{\sqrt 2} \frac{E_0}{\omega} 
    e^{ik \hat z}
    \left(
    e^{-i\omega t}\hat \epsilon_\pm
    +
    e^{i\omega t}\hat \epsilon_\mp
    \right)
    -
    \frac{B_0}{2} \boldsymbol r\times \hat{e}_z 
\end{align}

The minimal coupling Hamiltonian $H_0$ can be decomposed into Bohr's hydrogenic Hamiltonian, the linear Zeeman shifts, the diamagnetic term, and the E1 transition.

\begin{align}
    H_0 
    &=
    \stackrel{\text{Bohr atom}}{
        \frac{\boldsymbol{p}^2}{2m} - \frac{Ze^2}{4\pi\varepsilon_0}\frac{1}{|\boldsymbol r|}
    }
    +
    \stackrel{\text{Zeeman}}{
        \frac{e}{m}\boldsymbol{L}\cdot \boldsymbol{B}_0 + 
        \frac{e\hbar}{2m}\boldsymbol{\sigma}\cdot \boldsymbol{B}_0
    }
    \nonumber
    \\
    &\ \ \ 
    +
    \stackrel{\text{Diamagnetism}}{
        \frac{e^2}{12 m}\boldsymbol{B}^2_0|\boldsymbol{r}|^2
        \left(
            1 - \sqrt{\frac{4\pi}{5}}Y_{20}
        \right)
    }
    +
    \stackrel{\text{E1 transitions}}{
        \frac{1}{2}\boldsymbol{E_0}\cdot e\boldsymbol{r}
    }
    \label{eq:minimum_coupling_Hamiltonian}
\end{align}

We have taken the dipole approximation $e^{ikz}\sim 1$ and neglected the spin M1 transition. The fine-structure Hamiltonian can be decomposed into the familiar relativistic kinetic correction, the Darwin term, the spin-orbit coupling. The traditionally neglected angular magnetoelectric (AME) term, which couples the electron spin to the optical angular momentum $\propto(\boldsymbol{E}\times \boldsymbol{A})$, must also be accounted for at this order \cite{Mondal2015}.

\begin{align}
    H_\text{FS}
    &=
    \stackrel{\text{Relativistic kinetic}}{
        -\frac{\boldsymbol{p}^4}{8m^3c^2}
        + 
        \mathcal{O}\left(\frac{\mu_B B_0}{mc^2}\right)
    }
    \nonumber
    \\
    &\ \ \ 
    + 
    \stackrel{\text{Darwin}}{
        \frac{e\hbar^2}{8m^2c^2}\nabla\cdot\boldsymbol E
    }
    % \nonumber
    % \\
    % &\ \ \ 
    + 
    \stackrel{\text{Spin-Orbit Coupling}}{
        \frac{\hbar}{2m^2c}
        \frac{Z\alpha}{|\boldsymbol r|^3}
        \boldsymbol{S}\cdot \boldsymbol{L}
    }
    \nonumber
    \\
    &\ \ \ 
    + 
    \stackrel{\text{Angular Magnetoelectric}}{
        \frac{e^2\hbar}{4m^2c^2}
        \sigma\cdot(\boldsymbol{E}\times \boldsymbol{A})
    }
\end{align}

This AME term can be decomposed further, to yield a spin-toroidal transition $\propto (\boldsymbol{r}\times\boldsymbol{\sigma})$.

\begin{align}
    H_\text{AME}
    &=
    \frac{e^2\hbar}{4m^2c^2}
    \sigma\cdot(\boldsymbol{E}\times \boldsymbol{A})
    \nonumber
    \\
    &=
    -
    \stackrel{\text{T1 transition}}{
        \frac{\mu_B^2}{ec}
        \frac{Z\alpha}{r^3}
        \frac{i}{\omega}
        \frac{ E_0}{\sqrt 2}
        e^{ikz}
        \left(
            e^{-i\omega t}\hat \epsilon_\pm
            +
            e^{i\omega t}\hat \epsilon_\mp
        \right)
        \cdot
        (\boldsymbol{r}\times \boldsymbol{\sigma})
    }
    \nonumber
    \\
    &\ \ \ \ \ \ 
    +
    \stackrel{\text{Relativistic corrections}}{
        \mathcal{O}\left(\frac{\mu_B B_0}{mc^2}\right)
    }
    \label{eq:Kuprov_spin_T1}
\end{align}

We neglect terms on the order of relativistic corrections to the Zeeman shift. To account for level-mixing by the diamagnetic term, we include typically neglected couplings between states of different radial orders for all terms (e.g., the Darwin contact term couples different $n\mathrm{S}$ orbitals). Additional field-related terms that arise in the quartic term in the kinetic operator $(\boldsymbol p + e\boldsymbol A)^4$ as well as the sixth-order Foldy-Wouthuysen terms are neglected, being highly suppressed in the low-energy limit. \par 

%%%%%%%%%%%%%%%%%%%%%%%%%%%%%%%%%%%%%%%%%%%%%%%

\begin{figure}[t!]
    \centering
    \includesvg[width=\linewidth]{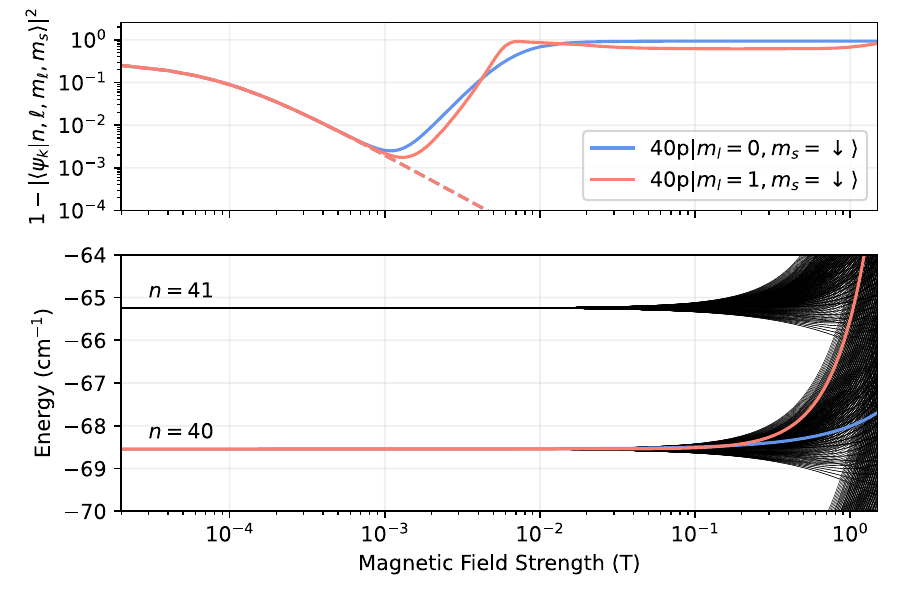}
    \caption{Upper: Inverse trace distance of the nearest eigenstate $\left|\psi\right\rangle$ with the decoupled basis $1 - T^2=1 - \left|\left\langle \psi| n',\ell,m_\ell,m_s \right\rangle\right|^2$ of the competing spin-orbit coupled states for spectroscopy. Without the diamagnetic term (dashed), the trace distance shrinks to arbitrarily small values. The diamagnetic mixing imposes a limit to spin-orbit decoupling, where the system transitions from the Paschen-Beck regime to the $\ell$-mixing regime (solid). Lower: Energy spectrum for target state $n' = 51$ up to $B = 0.5\ \text{T}$. At large magnetic fields, the energy spectra enters the $n$-mixing regime and exhibits level crossings between states of different $n$.}
    \label{fig:MagneticFieldDependence}
\end{figure}

\section{Toroidal Transitions in Magnetic Fields}
\label{sec:direct_observation}

Consider the dipole-allowed transitions from the ground state $n\text{S}\to n' {}\text{P}$ in hydrogenic atoms/ions. We start considering a $\sigma^+$ transition as shown in \autoref{fig:LevelDiagram}. Separation of the spin and orbit quantum numbers is required to observe the T1 transition against the E1 transition, which can be quantified by the trace distance of the closest eigenstate $\left|\psi\right\rangle$ to the uncoupled basis $1 - T^2=1 - \left|\left\langle \psi| n',\ell,m_\ell,m_s \right\rangle\right|^2$. For good visibility, we demand that $1 - T^2 \sim 10^{-11}$ is of the same order as the strength of the T1 transition against the E1 transition. \par

It was previously suggested that high-lying Rydberg states are more favourable for spectroscopy as the spin-orbit term scales as $\propto n^{-3}$. Without the diamagnetic term, the condition for good visibility of the T1 transition then appears at $B\sim 5\ \text T$ through the Zeeman shift  \cite{Kuprov2022}. \par

However, when the diamagnetism is taken into account, previous optimistic predictions for the experimentally-accessible conditions under which the T1 transition rate surpasses the E1 transition rate no longer exist. This term introduces couplings between states of the same parity ($\Delta \ell = 0, \pm 2$, see \autoref{eq:minimum_coupling_Hamiltonian}), breaking the spherical symmetry of the atom, and the energy level spectrum exhibit signatures of quantum chaos \cite{Friedrich1989}. In large magnetic fields, mixing occur between states of different angular momenta and $\ell$ is no longer a good quantum number ($\ell$-mixing regime). In even stronger fields, states of different radial indices begin to overlap, and $n$ is no longer a good quantum number ($n$-mixing regime) \cite{Gallagher1994}. \par

The diamagnetic term couples states with $\Delta \ell = 0, \pm 2$ and scales as $\propto n^{11}B^2$ \cite{Gallagher1994}. For high-lying Rydberg states, the diamagnetic term quickly dominates all other terms as the magnetic field increases. In addition, the degeneracy in each $n$-manifold scales as $n^2$, strongly diluting the dipole transition rates at large $n$ due to mixing between states of different angular momentum ($\ell$-mixing regime). The onset of the level-mixing regime (both $\ell$- and $n$-mixing) scales as $\propto  n^7B^2$ and limits the minimum $1 - T^2$ that can be realized \cite{Friedrich1989}. \par

We show this explicitly for $n'= 51$ in hydrogen. In \autoref{fig:MagneticFieldDependence}, the trace distance of the $51\text P$ states with the uncoupled basis asymptotically approaches zero without the diamagnetic term as the Zeeman shift suppresses the spin-orbit mixing (dotted). However, the inclusion of the diamagnetic term limits this due to l-mixing between terms of the different angular momentum, at approximately $0.4$ mT (solid). As the magnetic field increases, mixing between different n-manifolds dilute the eigenstates further, and at approximately $300$ mT, the $n=51$ manifold crosses the $n=50, 52$ manifolds (\autoref{fig:MagneticFieldDependence}, lower).

Both the T1 Hamiltonian and the spin-orbit term scales with the nuclear charge $\propto Z$. Ignoring the diamagnetic term (which has no dependence on $Z$), the scaling of the visibility of the T1 transition can be estimated as

\begin{equation}
    \propto\frac{H_\text{T1}}{H_\text{E1}}\times \frac{H_\text{Zeeman}}{H_\text{SO}}\sim Z^0
\end{equation}

where the dependence on $Z$ vanishes. This seems to rule out improvement of the visibility using highly-charged ions. For other alkali species or hydrogen-like atoms, the finite quantum defect introduces core-scattering that lowers the threshold for diamagnetic $\ell$- and $n$-mixing, worsening the prognosis for observation \cite{Gallagher1994, Courtney1995, Courtney_1996_LiChaos}. 

\begin{figure}[t!]
    \centering
    \includesvg[width=\linewidth]{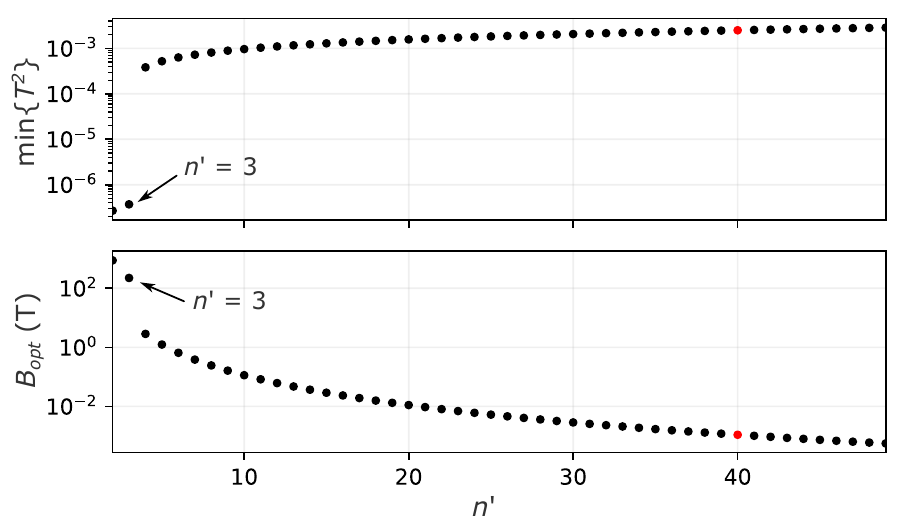}
    \caption{Trace distance of the closest eigenstate with the decoupled basis $1 - T^2$ and required magnetic field $B$ over $n'\text{P}$ states. The maximum decoupling occurs at low $n'$ for large magnetic fields. The discontinuity for the $n'=2,3$ states is due to the absence of the $\ell$-mixing regime at low $n$, which raises the threshold for diamagnetic mixing. The previously suggested $n'=51$ state is marked in green.}
    \label{fig:Optimal_LS_Separation}
\end{figure}

To avoid the contribution by the diamagnetic term $\propto n^{11}$, we instead suggest low-lying states may be more favourable for spectroscopy of the T1 transition. As shown in \autoref{fig:Optimal_LS_Separation}, for states with $n < 4$, the absence of the $F$-orbitals forbids $\ell$-mixing for the target $P$-orbitals, raising the threshold for diamagnetic mixing to the $n$-mixing regime. However, the magnetic fields required now approach experimentally-inaccessible values. We compute the best trace distance for up to $n = 70$ and the magnetic field required. \par

We find the optimal spin-orbit separation occurs for $n' = 2$ at $B \sim 886\ \text T$, with $1 - T^2\sim 2.7\times10^{-7}$. On the other hand, the previously suggested $n'=51$ Rydberg state finds an optimum at $1 - T^2\sim 2.9\times 10^{-3}$ at $B \sim 0.483\ \text{mT}$.  \par

\begin{figure}[t]
    \centering
    \includesvg[width=\linewidth]{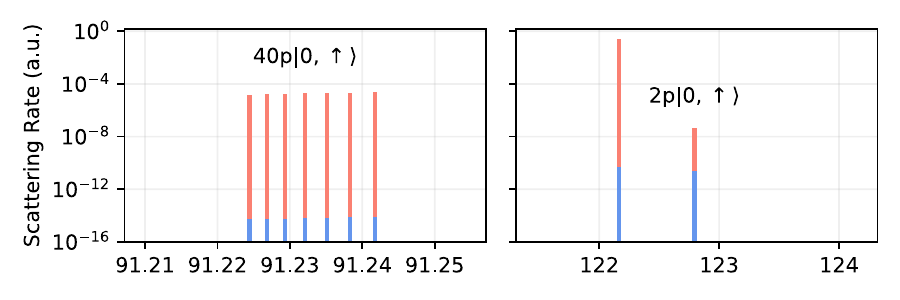}
    \caption{Spectroscopy of T1 scattering rates (relative to the dominant E1 transitions) of the $1\text{S}\to2\text{P}$ transition at $B=886\text{T}$ and the $1\text{S}\to51\text{P}$ at $B=4.83\times10^{-4}\text{T}$. The spectroscopic plots are centered at the probe transitions at $\lambda_{2\text{P}}=120.34795191\text{nm}$ and $\lambda_{51\text{P}}=91.210167421\text{nm}$ respectively.}
    \label{fig:spectroscopy_lines}
\end{figure}

Consider the dipole transition from the ground state $1 \mathrm{S} \to 2 \mathrm{P}$ and $1 \mathrm{S} \to 51\text{P}$ for hydrogen under illumination by a $\sigma^+$-polarized plane wave collinear with a static magnetic field (\autoref{fig:spectroscopy_lines}). The magnetic field is chosen to maximize the relative T1 scattering rate, which is estimated with Fermi's golden rule using \autoref{eq:minimum_coupling_Hamiltonian} and \autoref{eq:Kuprov_spin_T1}.  \par

At these magnetic fields, the two transitions mixed by spin-orbit coupling in the $2\text{P}$ orbital are strongly separated ($\Delta_{2\text{P}} \sim 2\pi\times12\ \text{THz}$) much larger than the estimated linewidth of the transition $\Gamma_{2\text{P}}\sim630\ \text{MHz}$. The toroidal term adds a small correction ($\sim 5.7\times10^{-4}$) to the on-resonant E1 scattering $1\text{S}|0,\downarrow\rangle\to2\text{P}|0,\uparrow\rangle$ at $\lambda_\text{2\text P}=120.34795191\ \text{nm}$, while the off-resonant E1 scattering from the other state $2\text{P}|1,\downarrow\rangle$ is suppressed at a factor of $\propto\left|\frac{\Gamma}{\Delta}\right|^2\sim3\times 10^{-9}$. \par

On the other hand, the two spin-orbit coupled states in the $51\text{P}$ orbital are less separated ($\Delta_{2\text{P}} \sim 2\pi\times7\ \text{MHz}$) with an estimated linewidth of $\Gamma_{51\text{P}}\sim31\ \text{kHz}$. The toroidal correction is much smaller ($\sim 7.8\times10^{-8}$) at the corresponding transition at $\lambda_\text{2\text P}=91.210167421\ \text{nm}$, while the off-resonant E1 scattering is larger $\propto\left|\frac{\Gamma}{\Delta}\right|^2\sim2\times10^{-5}$.
    
While the $n'=2$ state is still above the threshold for direct spectroscopy with clear visibility by four orders-of-magnitude, it is comparable to realized experimental tests of fundamental physics. For instance, the PNC amplitude in Caesium modifies the Stark-mixed electric dipole transition by $1$-in-$10^5$ in \cite{Wood1997}. Similar to the PNC detection approach, the weak T1 signal can be extracted by differential measurements taking advantage of the difference in temporal symmetry of the E1 and T1 transitions. For instance, for $S\rightarrow P$ transitions in hydrogenic atoms, the sign of T1 coupling term changes, inverting the direction of the static magnetic field, the light polarization, and the electronic spin state, while the E1 coupling remains unchanged. Hence, the far-field atomic electric and toroidal dipole emissions overlap and can be either in phase or out of phase \cite{Zagoskin2015}. The differential fluorescence containing then only the T1 contribution. We note that the required stability of magnetic field and laser polarization as well as the initial state fidelity is expected to be challenging to attain with such a differential measurement. \par

As suggested in ref. \cite{Kuprov2022}, the search of toroidal excitation is challenging in hydrogen atoms as the transition lies in the short UV spectral region where lasers are not available. Lithium atoms might be the best alternative as the fine structure energy shift is the smallest within the alkali family and the lowest $2S\rightarrow2P$ transition is conveniently located in the red spectral region. In addition, the odd-parity diamagnetic behavior of Lithium $P$-orbitals are nearly hydrogenic due to their naturally small quantum defect, which will have a small impact on the worsening of the visibility of the toroidal transition \cite{Gallagher1994, Courtney1995, Courtney_1996_LiChaos, Lithium_quantum_defects}.

%%%%%%%%%%%%%%%%%%%%%%%%%%%%%%%%%%%%%%%%%%%%%%%

\section{Conclusion}
\label{sec:conclusion}

Toroidal dipole excitations thus far are uncharted waters in atomic physics, and mapping this frontier may open the doors to new and exciting phenomena. While, observation of these transitions remain challenging -- the lowest Lyman line requires $B = 886\ \text{T}$ for a maximum overlap with the uncoupled basis states at $10^{-7}$ -- experimental observation is not entirely out of reach. Man-made magnetic fields of this magnitude are realizable in transient implosion-type set-ups \cite{Murakami2020} and current world records for steady-state magnetic fields only fall short by one order of magnitude \cite{Zhou2023}. Taking advantage of the different parity symmetry between E1 and T1 transitions, we propose differential measurement to remove the large E1 contribution. Another possibility for observation would be spectroscopic signals from astronomical objects known to have large magnetic fields \cite{Kaspi2017}. 

The observation and characterization of these transitions may present new opportunities in fields such as precision metrology \cite{Huntemann2012}, further tests of fundamental physics \cite{Zeldovich1957, Nanz2016}, or possibly applications in quantum computing as protected qubits \cite{Zagoskin2015}. Our work extends current literature towards designing experimentally realizable approaches to observe these transitions.
    
    %%%%%%%%%%%%%%%%%%%%%%%%%%%%%%%%%%%%%%%%%%%%%%%

\begin{acknowledgments}
    This work was supported by the Singapore Ministry of Education: Grant No MOE-T2EP50120-0005. The authors thank N. Zheludev and I. Kuprov for fruitful discussions. K. X. Lee. acknowledges support from the National Quantum Scholarship Scheme (NQSS). 
\end{acknowledgments}

\clearpage

\begin{widetext}
    \appendix
    
    \begin{appendices}
        \section{Equivalence of Toroidal Operators}
        \label{Appendix}
        
        Here, we show that the spin-toroidal dipole operator derived though the Foldy-Wouthuysen expansion by Kuprov \cite{Kuprov2022} and through the retardation expansion of the electromagnetic potentials by Marian \cite{Marian1996} are equivalent. Following \cite{Marian1996}, the atomic spin-toroidal dipole operator from the Pauli equation (in the non-relativistic limit) is given as
        
        \begin{equation}
            \boldsymbol{C}^s 
            =
            -\frac{1}{2} 
            \frac{Ze\hbar}{2mc}
            (\boldsymbol{r}\times \boldsymbol{\sigma})
        \end{equation}
        
        This operator couples to Maxwell's displacement current density $\boldsymbol{J}^d$,
        
        \begin{align}
            H_\text{T1}'
            =
            -\frac{4\pi}{c}
            \boldsymbol{C}^s\cdot \boldsymbol{J}^d
        \end{align}
        
        The displacement current can be obtained from the curl of the magnetic field, 
        
        \begin{align}
            \boldsymbol{J}^d(\boldsymbol{R},t)
            &=
            \frac{c}{4\pi}
            \boldsymbol{\nabla}\times \boldsymbol{B}
            \nonumber
            \\
            &=
            \frac{i\omega}{4\pi}
            \frac{E_0}{\sqrt 2}
            e^{ik\hat z}
            \left(
            e^{-i\omega t}\hat \epsilon_\pm
            +
            e^{i\omega t}\hat \epsilon_\mp
            \right)
        \end{align}
        
        The spin-toroidal Hamiltonian is thus given by
        
        \begin{align}
            H_\text{T1}'
            =
            -
            i\omega
            \frac{1}{2\sqrt 2}
            \frac{Z\mu_B E_0}{c^2}
            e^{ik \hat z}
            \left(
            e^{-i\omega t}\hat \epsilon_\pm
            +
            e^{i\omega t}\hat \epsilon_\mp
            \right)
            \cdot
            (\boldsymbol{r}\times \boldsymbol{\sigma})
        \end{align}
        
        We consider an on-resonant optical field for a dipole transition $\ket{n,\ell}\to\ket{n',\ell\pm1}$, such that the laser frequency $\omega$ can be given by the Rydberg relation \cite{Gallagher1994},
        
        \begin{equation}
            E_{n\to n'}
            =
            \hbar\omega
            =
            \frac{me^4}{2\hbar^2(4\pi\epsilon_0)^2}
            \left(\frac{1}{n^2} - \frac{1}{n'^2}\right),
            \ \ \ \ \ \ 
            n \neq n'
        \end{equation}
        
        Enforcing angular momentum selection rules, the only relevant matrix elements are likewise between $\ket{n,\ell}\to\ket{n',\ell\pm1}$. We take the following relation between dimensionless radial matrix elements, valid for $n\neq n',\ell'=\ell\pm1$ \cite{Blaive_2009}.
        
        \begin{equation}
            \left\langle n'l'\left|
                \left(\frac{r}{a_0}\right)^{-2}
            \right|nl\right\rangle
            =
            \frac14
            \left(
                \frac{1}{n^2}
                -
                \frac{1}{n'^2}
            \right)^2
            \left\langle n'l'\left|
                \left(\frac{r}{a_0}\right)
            \right|nl\right\rangle
        \end{equation}
        
        The Hamiltonian can then be recast into 
        
        %%%%% Abridged version
        \begin{align}
            H'_\text{T1}
            &=
            -
            i\omega
            \frac{1}{2\sqrt 2}
            \frac{Z\mu_B E_0}{c^2}
            e^{ik \hat z}
            \left(
                e^{-i\omega t}\hat \epsilon_\pm
                +
                e^{i\omega t}\hat \epsilon_\mp
            \right)
            \cdot
            (\boldsymbol{r}\times \boldsymbol{\sigma})
            \nonumber
            \\
            &=
            -
            \left(
                \frac{me^4}{2\hbar^3(4\pi\epsilon_0)^2}
                \left(\frac{1}{n^2} - \frac{1}{n'^2}\right)
            \right)^2
            \frac{i}{\omega}
            \frac{1}{2\sqrt 2}
            \frac{Z\mu_B E_0}{c^2}
            e^{ik \hat z}
            \left(
                e^{-i\omega t}\hat \epsilon_\pm
                +
                e^{i\omega t}\hat \epsilon_\mp
            \right)
            \cdot
            (\boldsymbol{r}\times \boldsymbol{\sigma})
            \nonumber
            \\
            &=
            -
            \frac{\mu_B^2}{ec}
            \frac{Z\alpha}{r^3}
            \frac{i}{\omega}
            \frac{ E_0}{\sqrt 2}
            e^{ik \hat z}
            \left(
                e^{-i\omega t}\hat \epsilon_\pm
                +
                e^{i\omega t}\hat \epsilon_\mp
            \right)
            \cdot
            (\boldsymbol{r}\times \boldsymbol{\sigma})
        \end{align}

        which returns us to the spin-toroidal Hamiltonian given in \autoref{eq:Kuprov_spin_T1} following Kuprov \cite{Kuprov2022}. 
    \end{appendices}
\end{widetext}

%%%%%%%%%%%%%%%%%%%%%%%%%%%%%%%%%%%%%%%%%%%%
\clearpage
\bibliographystyle{unsrt}
\bibliography{bibliography.bib}

\end{document}